\documentclass{article}
\usepackage{spconf,amsmath,graphicx}

\usepackage{enumitem}
\usepackage{threeparttable}
\usepackage{multirow}
\usepackage{multicol}
\usepackage{booktabs}
\usepackage{algorithm}
\usepackage{algorithmic}
\usepackage[dvipsnames]{xcolor}
\usepackage{cite}

\newcommand\fsenc[1][]{fast-slow encoder based transducer }
\newcommand\fsencend[1][]{fast-slow encoder based transducer}
\newcommand\Fsenc[1][]{Fast-slow encoder based transducer }

\title{Improving Fast-slow Encoder based Transducer\\with Streaming Deliberation}
\name{Ke Li, Jay Mahadeokar, Jinxi Guo, Yangyang Shi, Gil Keren, Ozlem Kalinli, Michael L. Seltzer, Duc Le}
\address{Meta AI, USA}

\begin{document}
\ninept

\maketitle
\begin{abstract}
This paper introduces a \fsenc with streaming deliberation for end-to-end automatic speech recognition. We aim to improve the recognition accuracy of the \fsenc while keeping its latency low by integrating a streaming deliberation model. Specifically, the deliberation model leverages partial hypotheses from the streaming fast encoder and implicitly learns to correct recognition errors. We modify the parallel beam search algorithm for \fsenc to be efficient and compatible with the deliberation model. In addition, the deliberation model is designed to process streaming data. To further improve the deliberation performance, a simple text augmentation approach is explored. We also compare LSTM and Conformer models for encoding partial hypotheses. Experiments on Librispeech and in-house data show relative WER reductions (WERRs) from 3\% to 5\% with a slight increase in model size and negligible extra token emission latency compared with \fsencend. Compared with vanilla neural transducers, the proposed deliberation model together with \fsenc obtains relative 10-11\% WERRs on Librispeech and around relative 6\% WERR on in-house data with smaller emission delays. 
\end{abstract}
\begin{keywords}
Fast-slow encoder-based transducer, Deliberation model, Parallel beam search, RNN-T, Conformer
\end{keywords}
\section{Introduction}
\label{sec:intro}
Low latency is a general requirement for voice assistants for good user experience. To this end, the streaming capability of an automatic speech recognition (ASR) system is a critical consideration. The recurrent neural network transducer (RNN-T)~\cite{graves2012sequence, rao2017exploring, he2019streaming, mahadeokar2021alignment}, which intrinsically supports streaming transcription, is a commonly adopted model for end-to-end (E2E) ASR~\cite{graves2012sequence, rao2017exploring, he2019streaming, mahadeokar2021alignment, graves2014towards, watanabe2017hybrid, amodei2016deep, chan2016listen, li2020comparison}.

Streaming RNN-Ts limit the lookahead context access of the input audio for latency control. Various two-pass approaches have been introduced to address the accuracy loss due to limited context~\cite{sainath2019two, Hu20deliberation,li2019integrating}. For example, an attention-based encoder-decoder model was used to rescore N-best hypotheses generated by an RNN-T in the 1st-pass~\cite{sainath2019two}. The deliberation model~\cite{Hu20deliberation} subsequently extended the previous work by adding a spelling correction component~\cite{Guo19spellingcorrection}. Another way to improve the performance of streaming RNN-Ts is to introduce a non-causal encoder with larger context. Narayanan et al.~\cite{narayanan2021cascaded} proposed 
a cascaded encoder-based RNN-T with a causal/streaming encoder and a non-causal/non-streaming encoder.
Li et al.~\cite{li2021better} applied two-pass beam search with cascaded encoders, where the 1st-pass uses a streaming encoder and the 2nd-pass uses a non-streaming encoder with left and right context. 


The non-streaming encoder in the cascaded architecture, however, introduces non-negligible latency. For latency-constrained use case, Mahadeokar et al.~\cite{mahadeokar2022streaming} proposed a streaming \fsenc design with both encoders having limited future lookahead. 
This streaming design achieves substantial accuracy improvement by the slow encoder, which takes multiple segments' output of the fast encoder as the input.
Mahadeokar et al.~\cite{mahadeokar2022streaming} also proposed a novel streaming parallel beam search for the fast-slow encoders where the search space for fast and slow encoders is shared. 

However, the accuracy of the \fsenc~\cite{mahadeokar2022streaming} is still compromised since the lookahead context of the streaming slow encoder is limited. In order to improve its accuracy while keeping the low latency, we integrate a streaming deliberation model into \fsenc that leverages partial hypotheses decoded from the fast encoder. 
Specifically, the partial hypotheses are passed through a text encoder. The resulting text embeddings are combined with the acoustic embeddings from the slow encoder by a merge module. The combined embeddings are then fed to the joiner. 
To allow the deliberation model to process streaming data, we limit the maximum length of partial hypotheses for the text encoder. During training, random masks are applied to partial hypotheses as a simple augmentation approach. 
Unlike~\cite{hu2022transducer} where an extra joiner is used, we use a shared joiner initialized by the joiner in a \fsencend. This simplifies the model design and makes training converge faster. We integrate the deliberation model into the parallel beam search algorithm. During inference, partial hypotheses are only generated by beam search from the fast encoder that is immediately before the call of the slow encoder beam search. 
Furthermore, we compare different text encoder architectures in terms of accuracy and emission delay, and discuss limitations of the deliberation approach. 



\section{Methods}
\vspace{-0.5em}
\label{sec:model}
\subsection{Fast-Slow Encoder Based Transducer}
We first introduce the baseline model, the streaming \fsenc~\cite{mahadeokar2022streaming}.
Compared to RNN-Ts with a single encoder, \fsenc improves recognition accuracy by leveraging more acoustic information through the slow encoder, which effectively has $K$ times acoustic context of the fast encoder. 
Different from the causal and non-causal cascaded encoders approach~\cite{narayanan2021cascaded}, both fast and slow encoders in~\cite{mahadeokar2022streaming} are streaming and has limited lookahead context for latency control. A novel parallel beam search algorithm was developed for \fsencend, where the slow encoder can update hypotheses decoded from the fast encoder and the search space for the two encoders is shared. In summary, the \fsenc achieves much better recognition accuracy without much increase in token emission delay compared to a baseline streaming RNN-T with similar architecture.
In this work, we build the proposed streaming deliberation model on top of the \fsencend, where the fast encoder is used to generate partial hypotheses for the streaming deliberation model.

\subsection{Streaming Deliberation Model for Fast-Slow Encoder Based Transducer}
In this section, we introduce the \fsenc with streaming deliberation including model details, training approach, and a simple text augmentation approach. 
\subsubsection{Model}
\begin{figure}[t]
    \centering
    \includegraphics[width=0.85\linewidth]{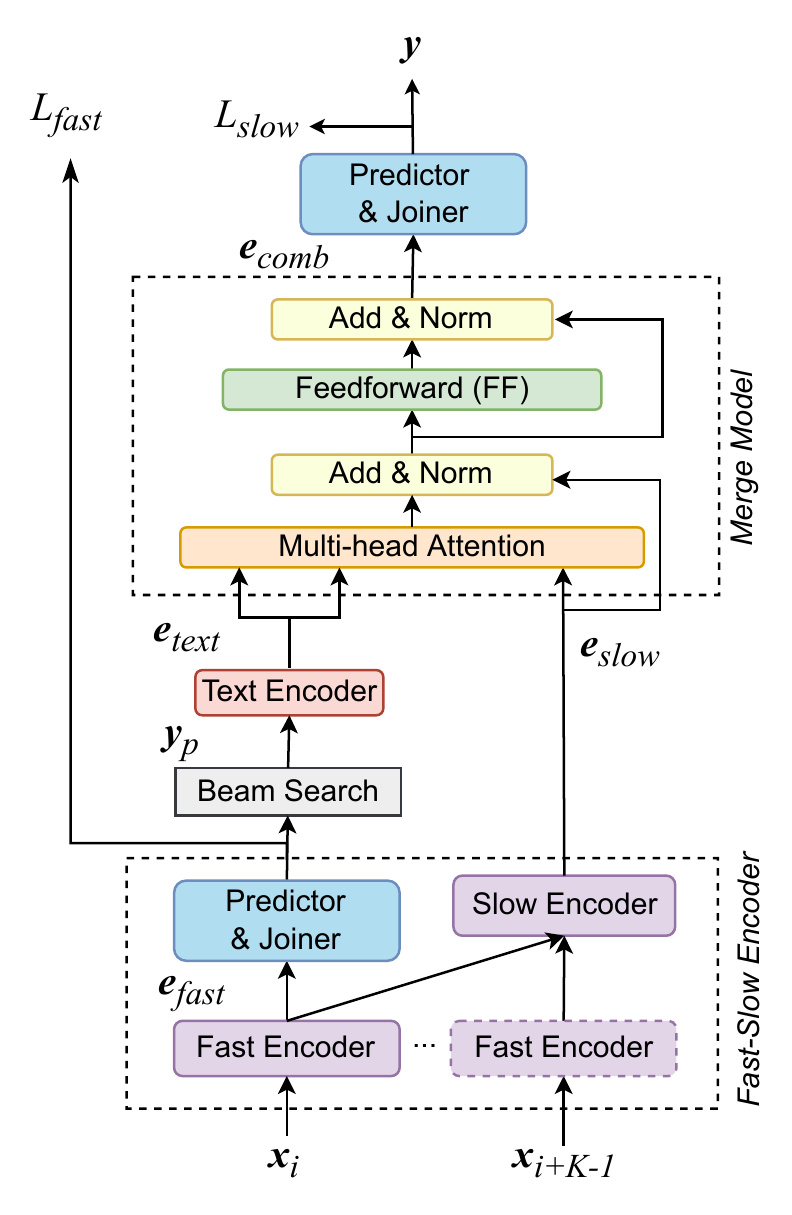}
    \vspace{-1em}
    \caption{\Fsenc with the streaming deliberation model. 
    }
    \vspace{-0.5em}
    \label{fig:fast-slow-deli}
\end{figure}

The deliberation model improves the recognition accuracy of \fsenc by leveraging partial decoded hypotheses from the fast encoder. The model is illustrated in Fig.~\ref{fig:fast-slow-deli}. 
The \fsenc is shown at the bottom of Fig.~\ref{fig:fast-slow-deli}. To support streaming processing, input acoustic frames are segmented into chunks and each chunk contains a certain number of audio frames. Each processed audio chunk $\mathbf{x}_{i}$ is concatenated with a small lookahead context and then fed into the fast encoder. The slow encoder is cascaded with the fast encoder and takes around $K$ outputs of the fast encoder as the input, e.g., $K$ can be two or more, as shown in Fig.~\ref{fig:fast-slow-deli}. 
Both the fast and slow encoders take a lookahead context size 1, which is effectively 40ms.  

We perform decoding on training data with the fast encoder to generate 1-best or N-best partial hypotheses. Let us denote 1-best hypothesis as $\mathbf{y}_{p}$. It is then encoded by a text encoder, which consists of a lookup embedding matrix and a neural network-based text encoder. 
The text embeddings $\mathbf{e}_{text}$ from the text encoder are combined with the acoustic embedding $\mathbf{e}_{slow}$ from the slow encoder by the `Merge Model' in Fig.~\ref{fig:fast-slow-deli}. The `Merge Model' consists of multiple blocks, where each block has a multi-head attention module and a feedforward (FF) network.
The acoustic embedding $\mathbf{e}_{slow}$ is the query, and the text embeddings $\mathbf{e}_{text}$ are key and value in the attention module. The residual connection adds a summarized text embedding with each acoustic embedding. The combined embedding from the `Merge Model' denoted as $\mathbf{e}_{comb}$ is then fed to the joiner along with the output of the predictor, which is omitted in Fig.~\ref{fig:fast-slow-deli} for simplicity. The predictor and joiner of the deliberation model share their weights with the predictor and joiner that connects to the fast encoder, respectively. 
To further reduce the model size, we also share the token embedding matrices between the text encoder and the predictor.

We compare the use of LSTM and Conformer architectures for the text encoder. To support streaming processing, the LSTM network has to be unidirectional.
For Conformer~\cite{Gulati2020ConformerCT}, we set a limit on the maximum number of tokens a partial hypothesis can contain; we found that this truncation with maximum length 20 has negligible impact on the word error rate (WER).
Our implementation of the streaming deliberation model is different from~\cite{hu2022transducer} considering the slow encoder beam search updates the output of the fast encoder so partial decoded hypotheses from the fast encoder vary over time.
The partial hypotheses are passed through the text encoder every time before the beam search call over the slow encoder. Since the number of beam search calls over the slow encoder are much fewer than the number of beam search calls over the fast encoder, the extra computation cost of the deliberation model is not much. 

\subsubsection{Joint Training}
The training objective of the proposed fast-slow encoders with a streaming deliberation module is
\begin{align}
L = L_{\text{slow}} + \lambda L_{\text{fast}}
\label{eq:loss}
\end{align}
where $0<\lambda < 1$. Both $L_\text{fast}$ and $L_\text{slow}$ use alignment restricted RNN-T loss~\cite{mahadeokar2021alignment}. As shown in Fig.~\ref{fig:fast-slow-deli}, $L_\text{fast}$ is computed from the fast encoder and $L_\text{slow}$ is computed from the slow encoder with deliberation module. Different from previous work ~\cite{narayanan2021cascaded} where the causal encoder and the non-causal encoder randomly use different training samples within a minibatch, both fast and slow encoders in our model use all the training data. 

The training procedure has two steps:
\begin{itemize}
    \item Train the \fsenc as in~\cite{mahadeokar2021alignment}.
    \item Jointly train the \fsenc initialized from the first step with the deliberation model.
\end{itemize}
In the second training step, the \fsenc is jointly optimized with the deliberation model. We experimented with freezing all or part of \fsenc model parameters but neither outperformed joint training.

Inspired by~\cite{Hu2022ImprovingDB}, we experiment with a simple text augmentation method to increase the diversity of the partial hypotheses for deliberation. We randomly mask out word piece tokens in partial hypotheses with a small probability during training. The mask token we use is the blank token, so there is no need to change the vocabulary. This is different from alignment augmentation~\cite{wang2022deliberation} where the blank token is already included in decoded hypotheses thus a new symbol that represents the mask token is required.  
Similar to~\cite{Hu2022ImprovingDB}, this random masking trick is applied only in training as an augmentation approach, thus no change is required in inference. 

\subsection{Parallel Beam Search with Deliberation Model}
We modify the original parallel beam search algorithm for the \fsenc~\cite{mahadeokar2022streaming} to support the streaming deliberation model. 
The modified algorithm is described below (modifications are in {\color{Plum}purple}).
\begin{algorithm}[ht] 
\caption{Parallel beam search for fast-slow encoders with streaming deliberation.} 
\label{parallel_beamsearch_algo} 
\begin{algorithmic} 
\STATE $T^{\mathrm{f}} =  \mathtt{fastSegmentSize}()$ 
\STATE $T^{\mathrm{s}} = \mathtt{slowSegmentSize}()$
\STATE $N^{\mathrm{f}} = \mathtt{fastBeamSize}()$
\STATE $N^{\mathrm{s}} = \mathtt{slowBeamSize}()$
\STATE $B^{\mathrm{f}} \gets \emptyset ; B^{\mathrm{s}} \gets \emptyset$
\STATE $(H^{\mathrm{f}}, H^{\mathrm{s}}) \gets \mathtt{initModelState}()$
\STATE $\Gamma \gets \mathtt{initSearchSpace}()$
\STATE $I^{\mathrm{s}} \gets \emptyset$
\FOR {$t = T^{\mathrm{f}}$ to $T$ by $T^{\mathrm{f}}$}
    \STATE $(I^{\mathrm{f}}, R^{\mathrm{f}}) = \mathtt{getFeatures}(t, t - T^{\mathrm{f}})$ 
    \STATE $(\mathbf{e}_{\mathrm{f}}, H^{\mathrm{f}}) = \mathtt{encodeFast}(I^{\mathrm{f}}, (I^{\mathrm{f}}, R^{\mathrm{f}}, H^{\mathrm{f}}))$
    \STATE $B^{\mathrm{f}} \gets \mathtt{beamSearch}(\mathbf{e}_{\mathrm{f}}, B^{\mathrm{f}}, N^{\mathrm{f}}, \Gamma, H^{\mathrm{f}})$
    \begin{*}
    \color{Plum}
    \IF{ $t \mod T^{\mathrm{s}} = T^{\mathrm{s}} - 1$}
        \STATE $\mathbf{y}_{p}$ = \text{getBestHypo}($B^{\mathrm{f}})$
    \ENDIF
    \end{*}
    \STATE $I^{\mathrm{s}} \gets \mathtt{concat}(I^{\mathrm{s}}, \mathbf{e}_{\mathrm{f}})$
    \IF{ $t \mod T^{\mathrm{s}} = 0$ or $t = T$}
        \STATE $(\mathbf{e}^{\mathrm{s}}, H^{\mathrm{s}}) = \mathtt{encodeSlow}(I^{\mathrm{s}}, (I^{\mathrm{s}}, R^{\mathrm{s}}, H^{\mathrm{s}}))$
        \begin{*}
        \color{Plum}
        \STATE $\mathbf{e}^{\mathrm{comb}} = \mathtt{encodeDeliberation}(\mathbf{e}_{\mathrm{s}}, \mathbf{y}_{p})$
        \end{*}
        \STATE $B^{\mathrm{s}} \gets \mathtt{beamSearch}(
        \begin{*}
        \color{Plum}\mathbf{e}^{\mathrm{comb}}
        \end{*}, B^{\mathrm{s}}, N^{\mathrm{s}}, \Gamma, H^{\mathrm{s}})$
        \STATE $I^{\mathrm{s}} \gets \emptyset$
        \STATE $B^{\mathrm{f}} \gets B^{\mathrm{s}}$
    \ENDIF
\ENDFOR
\RETURN  $y$ with highest $\log Pr(y) / |y| $ in $B^{\mathrm{s}}$ 
\end{algorithmic}
\end{algorithm}

Note that `f' denotes `fast' and `s' denotes `slow' in the above algorithm. Let us assume $\Gamma$ represents the shared decoding space for fast and slow encoders. $T^{f}$ and $T^{s}$ denote chunk or segment size of fast and slow encoders and $T$ denotes acoustic sequence length. 
$N^{f}$ and $N^{s}$ denote beam sizes for decoding from the fast encoder and the slow encoder respectively. $B^{f}$ and $B^{s}$ represent the N-best hypotheses generated using fast-slow encoders, respectively. $H^{f}$ and $H^{s}$ denote decoding states for fast and slow encoders. The 1-best partial hypothesis from the fast encoder is denoted as $y_{p}$. 

The key process of the algorithm contains interval calls of beam search over fast and slow encoders. After initialization, we iterate over acoustic frames from time step 0 to $T$ in the interval of $T^{f}$ and run beam search with the fast encoder. The output acoustic embedding $\mathbf{e}^{f}$ from the fast encoder is accumulated until we need to perform beam search with the slow encoder. The accumulated embedding represented by $I^{s}$ is used as input for the slow encoder. After the beam search call of the slow encoder, we update $B^{f}$ with $B^{s}$ and run the next fast encoder beam search on the next chunk of acoustic frames until the end. 

The highlighted part in purple shows how deliberation is integrated into the algorithm. When we process the chunk of acoustic frames just before the next beam search call over the slow encoder, we trace the 1-best partial hypothesis $y_{p}$ from the current beam search over the fast encoder. The $\mathtt{encodeDeliberation}$ operation then encodes $y_{p}$ and combines it with acoustic embedding $\mathbf{e}^{s}$ from the slow encoder by the merge model. The resulting embedding $\mathbf{e}^{comb}$ is then used for beam search to generate $B^{f}$. Note that $\mathtt{encodeDeliberation}$ call happens infrequently, only after $\mathtt{encodeSlow}$ is called. This saves computation from the deliberation model. 

\section{Experimental Setup}
\label{sec:exps}
\subsection{Datasets}
\subsubsection{Librispeech}
\label{subsec:librispeech}
The Librispeech~\cite{panayotov2015librispeech} corpus contains 960 hours of labeled speech. 80-dimensional filter bank features are extracted from a 25 ms window with a stride of 10 ms. We apply SpecAugment~\cite{park2019specaugment} with mask parameter $F$ = 27, ten time masks with maximum time-mask ratio $p_\text{s}$ = 0.05, and speed perturbation. 

\subsubsection{Large-Scale In-House Data}
\label{subsec:in_house_data}
Our in-house training set combines two sources. The first consists of 20K hours of English video data publicly shared by Facebook users; all videos are completely de-identified before transcription. The second contains 20K hours of manually transcribed de-identified English data with no user-identifiable information in the voice assistant domain. All utterances are morphed when researchers manually access them to further de-identify the speaker. Note that the data is not morphed during training. We further augment the data with speed perturbation, simulated room impulse response, and background noise, resulting in 83M utterances (145K hours).

We evaluate our models on two in-house test sets:

\textbf{VA1} -- 10.2K hand-transcribed de-identified short-form utterances (less than five words on average) in the voice assistant domain, collected from internal volunteers. The participants consist of households that have agreed to have their voice activity reviewed and analyzed.

\textbf{VA2} -- 44.2K hand-transcribed de-identified short-form utterances in the voice assistant domain, collected by a third-party data vendor.


\subsection{Models}
Our baseline RNN-T models use 20 layers of Emformer~\cite{shi2021emformer} as the encoder. The joiner is a 1-layer feed-forward network, and the predictor is a 3-layer LSTM. The word piece vocabulary size is 5001 for Librispeech and 4096 for in-house data. The total number of parameters of the baseline RNN-T model is approximately 79M on Librispeech and 78M on in-house data.
Baseline \fsenc consists of 15 fast encoder layers and 5 slow encoder layers. Each encoder layer uses the Emformer architecture. 
We choose $\lambda$ = 0.5 for the fast encoder loss (Equation~\ref{eq:loss}). We experiment with both LSTM and Conformer for text encoders. For Conformers, we limit the total token size of each partial hypothesis up to 20. 
The Merge Model consists of 1-layer multi-head attention with one head and a feed-forward network.

We evaluate recognition accuracy using WER and measure token emission latency by tracking average, P95 and P99 emission delays. Emission delay is defined as the time from when the token is spoken to when the transcript of the token is emitted~\cite{mahadeokar2021alignment}. 

Parameter setups to train baseline RNN-T and \fsenc follow~\cite{mahadeokar2022streaming}. For training the streaming deliberation models, we use a 100 times smaller learning rate and half the number of total epochs compared to the baseline models. To improve training efficiency, we set the beam size to 1 when decoding training data. Compared to larger beam sizes, e.g, 5 or 10, beam size 1 significantly accelerates training without degrading model accuracy. 
During inference, we use beam size 10 for all models.

\section{Experimental Results}
\subsection{WERs and Latency on Librispeech}
We compare the WER and emission delay (ED) of the proposed model and baseline RNN-T and \fsenc on Librispeech. The total extra parameters from the deliberation module with a 3-layer Conformer text encoder is around 16M. We use context size 160ms for the fast encoder and 800ms for the slow encoder. As shown in Table~\ref{tab:librispeech}, the deliberation model with random mask probability 0.1 achieves 5\% and 3\% WERR on test-clean and test-other, respectively, compared to \fsencend. It slightly increases average emission delay while having no effect on P95 and P99 compared to fast-slow encoders. The relative WER reductions are 10-11\% compared to RNN-Ts. As for random masking, small yet consistent gains are achieved by using small probabilities such as 0.05 and 0.1. Using a masking probability greater than 0.2 does not show improvement.
\begin{table}[ht]
    \setlength{\tabcolsep}{2.0pt}
    \caption{WERs (\%) and ED (ms) from baseline RNN-Ts and proposed \fsenc on Librispeech.}
    \vspace{1em}
    \label{tab:librispeech}
    \centering
    \scalebox{1}{
    \begin{tabular}{ l c c c c c c c c}
    \toprule
        \multirow{2}{*}{model} & 
        \multirow{2}{*}{context} & \multicolumn{2}{c}{WER} & &
        \multicolumn{3}{c}{ED} & \\\cmidrule{3-4}
        \cmidrule {6-8} 
        &  & test-clean & test-other & & avg & P95 &P99 \\
        \midrule
        RNN-T & 160 & 3.54 & 8.85 && 344 & 480 & 560  \\
        fast-slow & \multirow{3}{*}{160/800} & 3.32 & 8.17 && 335 & 480 & 600\\
        \;\;+ deliberation &  & 3.20 & 7.98 && 337  & 480 & 600 \\
        \;\;\;\;+ mask p = 0.1 &  & \textbf{3.15} & \textbf{7.90} && 337  & 480 & 600 \\
        \bottomrule
    \end{tabular}}
    \vspace{-1em}
\end{table}

\subsection{Effect of Text Encoder Architecture}
Text encoder is a key component in the deliberation model and can affect both accuracy and latency. We experiment with both LSTM and Conformer architectures in 1-layer and 3-layer setups. All deliberation models are trained with a random masking probability of 0.1. Table \ref{tab:text_encoder} presents the comparison results on Librispeech. For 1-layer text encoders, LSTM performs slightly better than Conformer. This is expected since usually deeper Conformer architectures are more powerful. For the 3-layer setup, Conformer performs slightly better. There is no significant difference in ED for both architectures. 
\begin{table}[ht]
    \setlength{\tabcolsep}{1.5pt}
    \caption{WERs (\%) and ED (ms) from LSTM and Conformer text encoders of the deliberation model on Librispeech.}
    \vspace{1em}
    \label{tab:text_encoder}
    \centering
    \scalebox{1}{
    \begin{tabular}{ l  c  c  c c c c c}
    \toprule
        \multirow{2}{*}{model} & 
        \multirow{2}{*}{context} & \multicolumn{2}{c}{WER} &&
        \multicolumn{2}{c}{ED} & \\
        \cmidrule {3-4}\cmidrule{6-7} 
        &  & test-clean & test-other && avg & P99 \\
        \midrule
        RNN-T & 160 & 3.54 & 8.85 && 344 & 560  \\
        fast-slow & 160/800 & 3.32 & 8.17 && 335 & 600\\
        \midrule
        \;+ delib (LSTM 1-L)  & \multirow{2}{*}{160/800} & 3.17 & 7.96 && 337 & 600 \\
        \;+ delib (Conformer 1-L ) & & 3.25 & 8.01 && 337 & 600\\
        \midrule
        \;+ delib (LSTM 3-L) & \multirow{2}{*}{160/800} & 3.20 & 7.93 && 337 & 600\\
        \;+ delib (Conformer 3-L) & &3.15 & 7.90 && 337 & 600 \\
        \bottomrule
    \end{tabular}}
    \vspace{-1em}
\end{table}

\subsection {Limitation of Deliberation}
The key capacity of deliberation model is its potential correction ability learned by observing hypotheses with errors in training. Therefore, one key limitation of this approach is that it may not help much for short utterances without enough errors. To test this hypothesis, we conduct an analysis on the Librispeech dataset. We split each test set into two parts, one containing utterances shorter than 3s and the other with utterances longer than 3s. WERs in Table~\ref{tab:length_analysis} show improvements are mainly from longer utterances as hypothesized. This indicates the deliberation model is more powerful for dictation application.
\begin{table}[ht]
    \setlength{\tabcolsep}{2.0pt}
    \caption{WERs (\%) from test sets with different duration on Librispeech.}
    \vspace{1em}
    \label{tab:length_analysis}
    \centering
    \scalebox{1}{
    \begin{tabular}{ l c c c c c c}
    \toprule
        \multirow{2}{*}{model} & \multicolumn{2}{c}{test-clean} & & \multicolumn{2}{c}{test-other}\\\cmidrule{2-3}\cmidrule{5-6}
        & $>$3s & $\le$3s &&  $>$3s & $\le$3s \\
        \midrule
        fast-slow & 3.3 & 4.3 && 8.0 & 10.3\\
        \;\;+ deliberation & 3.1 & 4.4 && 7.7 & 10.3 \\
        \bottomrule
    \end{tabular}}
    \vspace{-1em}
\end{table}

\subsection{WERs and Latency on in-house Data}
This section contains the results of the proposed model on large-scale in-house data. We use a 3-layer Conformer as the text encoder with a dropout rate of 0.1. Compared to the fast-slow RNN-T baseline, the deliberation model obtains around 3\% WERR on `VA2' while no improvement on `VA1' where utterances have less than five words on average. The latter result is consistent with the analysis in the above section. Similar to Librispeech, on average, there is only a small extra emission delay introduced by the deliberation model. P95 emission delays are the same for all models we experiment with.
\begin{table}[ht]
    \setlength{\tabcolsep}{2.0pt}
    \caption{WERs (\%) and ED (ms) from baselines RNN-Ts and proposed \fsenc on in-house data.}
    \vspace{1em}
    \label{tab:in-house}
    \centering
    \scalebox{1}{
    \begin{tabular}{ l c c c c c c c c}
    \toprule
        \multirow{2}{*}{model} &
        \multirow{2}{*}{context} & \multicolumn{2}{c}{WER} & &
        \multicolumn{2}{c}{ED} & \\
        \cmidrule {3-4}\cmidrule{6-7} 
        &  & VA1 & VA2 && Avg & P95\\
        \midrule
        RNN-T & 160 & 4.73 & 12.89 & & 390 & 560 \\
        fast-slow & \multirow{3}{*}{160/800} & \textbf{4.65} & 12.43 && 369 & 560 \\
        \;\;+ deliberation & &  4.70 & 12.13 && 373  & 560 \\
        \;\;\;\;+ mask p = 0.1 & & 4.68& \textbf{12.08} && 373  & 560\\
        \bottomrule
    \end{tabular}}
    \vspace{-1em}
\end{table}

\section{Conclusion and Future Work}
This paper introduces a streaming deliberation model for \fsencend, which improves the recognition accuracy of \fsenc by leveraging partial hypotheses from the fast encoder. The proposed deliberation model shows 3-5\% WERR on both Librispeech and in-house data compared to \fsencend, and 10-11\% WERR on Librispeech and up to 6\% WERR on in-house data compared with baseline RNN-Ts. The proposed deliberation model introduces negligible extra emission delays. 
In the future, we will explore alternative text augmentation approaches for further accuracy improvement.

\vfill\pagebreak

\bibliographystyle{IEEEbib}
\bibliography{refs}

\begin{thebibliography}{10}

\bibitem{graves2012sequence}
Alex Graves,
\newblock ``Sequence transduction with recurrent neural networks,''
\newblock {\em arXiv preprint arXiv:1211.3711}, 2012.

\bibitem{rao2017exploring}
Kanishka Rao, Ha{\c{s}}im Sak, and Rohit Prabhavalkar,
\newblock ``Exploring architectures, data and units for streaming end-to-end
  speech recognition with {RNN}-transducer,''
\newblock in {\em Proc. of ASRU}, 2017.

\bibitem{he2019streaming}
Yanzhang He, Tara~N Sainath, Rohit Prabhavalkar, Ian McGraw, Raziel Alvarez,
  Ding Zhao, David Rybach, Anjuli Kannan, Yonghui Wu, Ruoming Pang, et~al.,
\newblock ``Streaming end-to-end speech recognition for mobile devices,''
\newblock in {\em Proc. of ICASSP}, 2019.

\bibitem{mahadeokar2021alignment}
Jay Mahadeokar, Yuan Shangguan, Duc Le, Gil Keren, Hang Su, Thong Le,
  Ching-Feng Yeh, Christian Fuegen, and Michael~L Seltzer,
\newblock ``Alignment restricted streaming recurrent neural network
  transducer,''
\newblock in {\em Proc. of SLT}, 2021.

\bibitem{graves2014towards}
Alex Graves and Navdeep Jaitly,
\newblock ``Towards end-to-end speech recognition with recurrent neural
  networks,''
\newblock in {\em Proc. of ICML}, 2014.

\bibitem{watanabe2017hybrid}
Shinji Watanabe, Takaaki Hori, Suyoun Kim, John~R Hershey, and Tomoki Hayashi,
\newblock ``Hybrid {CTC}/attention architecture for end-to-end speech
  recognition,''
\newblock {\em IEEE Journal of Selected Topics in Signal Processing}, 2017.

\bibitem{amodei2016deep}
Dario Amodei, Sundaram Ananthanarayanan, Rishita Anubhai, Jingliang Bai, Eric
  Battenberg, Carl Case, Jared Casper, Bryan Catanzaro, Qiang Cheng, Guoliang
  Chen, et~al.,
\newblock ``Deep speech 2: End-to-end speech recognition in {E}nglish and
  {M}andarin,''
\newblock in {\em Proc. of ICML}, 2016.

\bibitem{chan2016listen}
William Chan, Navdeep Jaitly, Quoc Le, and Oriol Vinyals,
\newblock ``Listen, attend and spell: A neural network for large vocabulary
  conversational speech recognition,''
\newblock in {\em Proc. of ICASSP}, 2016.

\bibitem{li2020comparison}
Jinyu Li, Yu~Wu, Yashesh Gaur, Chengyi Wang, Rui Zhao, and Shujie Liu,
\newblock ``On the comparison of popular end-to-end models for large scale
  speech recognition,''
\newblock in {\em Proc. of Interspeech}, 2020.

\bibitem{sainath2019two}
Tara~N Sainath, Ruoming Pang, David Rybach, Yanzhang He, Rohit Prabhavalkar,
  Wei Li, Mirk{\'o} Visontai, Qiao Liang, Trevor Strohman, Yonghui Wu, et~al.,
\newblock ``Two-pass end-to-end speech recognition,''
\newblock {\em arXiv preprint arXiv:1908.10992}, 2019.

\bibitem{Hu20deliberation}
Kevin Hu, Rohit Prabhavalkar, Ruoming Pang, and Tara Sainath,
\newblock ``Deliberation model based two-pass end-to-end speech recognition,''
\newblock in {\em Proc. of ICASSP}, 2020.

\bibitem{li2019integrating}
Qiujia Li, Chao Zhang, and Philip~C Woodland,
\newblock ``Integrating source-channel and attention-based sequence-to-sequence
  models for speech recognition,''
\newblock in {\em Proc. of ASRU}, 2019.

\bibitem{Guo19spellingcorrection}
Jinxi Guo, Tara Sainath, and Ron Weiss,
\newblock ``A spelling correction model for end-to-end speech recognition,''
\newblock in {\em Proc. of ICASSP}, 2019.

\bibitem{narayanan2021cascaded}
Arun Narayanan, Tara~N Sainath, Ruoming Pang, Jiahui Yu, Chung-Cheng Chiu,
  Rohit Prabhavalkar, Ehsan Variani, and Trevor Strohman,
\newblock ``Cascaded encoders for unifying streaming and non-streaming {ASR},''
\newblock in {\em Proc. of ICASSP}, 2021.

\bibitem{li2021better}
Bo~Li, Anmol Gulati, Jiahui Yu, Tara~N Sainath, Chung-Cheng Chiu, Arun
  Narayanan, Shuo-Yiin Chang, Ruoming Pang, Yanzhang He, James Qin, et~al.,
\newblock ``A better and faster end-to-end model for streaming {ASR},''
\newblock in {\em Proc. of ICASSP}, 2021.

\bibitem{mahadeokar2022streaming}
Jay Mahadeokar, Yangyang Shi, Ke~Li, Duc Le, Jiedan Zhu, Vikas Chandra, Ozlem
  Kalinli, and Michael~L. Seltzer,
\newblock ``Streaming parallel transducer beam search with fast-slow cascaded
  encoders,''
\newblock in {\em Proc. of Interspeech}, 2022.

\bibitem{hu2022transducer}
Ke~Hu, Tara~N Sainath, Arun Narayanan, Ruoming Pang, and Trevor Strohman,
\newblock ``Transducer-based streaming deliberation for cascaded encoders,''
\newblock in {\em Proc. of ICASSP}, 2022.

\bibitem{Gulati2020ConformerCT}
Anmol Gulati, James Qin, Chung-Cheng Chiu, Niki Parmar, Yu~Zhang, Jiahui Yu,
  Wei Han, Shibo Wang, Zhengdong Zhang, Yonghui Wu, and Ruoming Pang,
\newblock ``Conformer: Convolution-augmented transformer for speech
  recognition,''
\newblock in {\em Proc. of Interspeech}, 2020.

\bibitem{Hu2022ImprovingDB}
Ke~Hu, Tara~N. Sainath, Yanzhang He, Rohit Prabhavalkar, Trevor Strohman,
  Sepand Mavandadi, and Weiran Wang,
\newblock ``Improving deliberation by text-only and semi-supervised training,''
\newblock in {\em Proc. of Interspeech}, 2022.

\bibitem{wang2022deliberation}
Weiran Wang, Ke~Hu, and Tara~N Sainath,
\newblock ``Deliberation of streaming {RNN}-transducer by non-autoregressive
  decoding,''
\newblock in {\em Proc. of ICASSP}, 2022.

\bibitem{panayotov2015librispeech}
Vassil Panayotov, Guoguo Chen, Daniel Povey, and Sanjeev Khudanpur,
\newblock ``Librispeech: an {ASR} corpus based on public domain audio books,''
\newblock in {\em Proc. of ICASSP}, 2015.

\bibitem{park2019specaugment}
Daniel~S Park, William Chan, Yu~Zhang, Chung-Cheng Chiu, Barret Zoph, Ekin~D
  Cubuk, and Quoc~V Le,
\newblock ``Spec{A}ugment: A simple data augmentation method for automatic
  speech recognition,''
\newblock {\em arXiv preprint arXiv:1904.08779}, 2019.

\bibitem{shi2021emformer}
Yangyang Shi, Yongqiang Wang, Chunyang Wu, Ching-Feng Yeh, Julian Chan, Frank
  Zhang, Duc Le, and Mike Seltzer,
\newblock ``Emformer: Efficient memory transformer based acoustic model for low
  latency streaming speech recognition,''
\newblock in {\em Proc. of ICASSP}, 2021.

\end{thebibliography}

\end{document}